\documentclass{PoS}

\usepackage{alltt}
\usepackage{axodraw}
\newcommand{\ep}{\varepsilon}
\newcommand{\ovl}{\relax}

\newcommand{\beq}{\begin{equation}}
\newcommand{\eeq}{\end{equation}}
\newcommand{\ice}[1]{\relax}

\title{Applications of FIESTA}
\ShortTitle{Applications of FIESTA}

\author{\speaker{M.~Tentyukov}\\
        Institut f\"ur Theoretische Teilchenphysik, Karlsruhe
        Institute of Technology (KIT), D-76128 Karlsruhe, Germany\\
        E-mail: \email{tentukov@particle.uni-karlsruhe.de}}
\author{A.V.~Smirnov\\
        Scientific Research Computing Center, Moscow State
        University, 119992 Moscow, Russia\\
        E-mail: \email{asmirnov80@gmail.com}}

\abstract{
  Sector decomposition in its practical aspect is a constructive
  method used to evaluate Feynman integrals numerically. We present a
  new program performing the sector decomposition and integrating the
  expression afterwards. The program can be also used in order to
  expand Feynman integrals automatically in limits of momenta and
  masses with the use of sector decompositions and Mellin--Barnes
  representations. The program is parallelizable on modern multicore
  computers and even on multiple computers. Also we demonstrate some
  new numerical results for four-loop massless propagator master integrals.
}

\FullConference{13th International Workshop on Advanced Computing and
  Analysis Techniques in Physics Research\\
                 February 22-27, 2010\\
                 Jaipur, India}

\begin{document}

\section{Introduction}

Originally sector decomposition was used as a tool for
analyzing the convergence and proving theorems on renormalization and
asymptotic expansions of Feynman integrals
\cite{Hepp,theory,BM,BdCMPo,books1a}.
After \cite{BH}, the 
sector decomposition approach has become an efficient tool for numerical evaluating Feynman
integrals (see Ref.~\cite{Heinrich} for a recent review).
At present, there are two public codes performing the
sector decomposition \cite{BognerWeinzierl} and \cite{FIESTA}. 
%
%available, one is a code by Bogner and Weinzierl
%\cite{BognerWeinzierl} and the other one by Smirnov, Smirnov,
%Tentyukov \cite{FIESTA}. 
%
The latter one was developed by the present authors; it is 
named {\tt FIESTA} which stands for ``Feynman Integral
Evaluation by a Sector decomposiTion Approach''. Recently 
{\tt FIESTA} was greatly improved in various aspects
\cite{FIESTA2}.

During the last year {\tt FIESTA} was widely used,
some of application are listed in \cite{FIESTA-appl}. In \cite{propagators} we
used {\tt FIESTA} in order to confirm numerically the recent analytic
results for {\it master integrals} (MI's) for four-loop massless 
propagators which recently were analytically evaluated in
\cite{BaCh}. Here we provide some more numerical results for extra
orders in epsilon expansions for these MI's.

\section{Theoretical background and software structure}

{\tt FIESTA} calculates Feynman integrals with the sector decomposition approach.
%It is based on the $\alpha$-representation of Feynman integrals.
%
After performing Dirac and Lorentz algebra one is left with a scalar dimensionally regularized Feynman integral \cite{dimreg}
%\begin{eqnarray}
  $F(a_1,\ldots,a_n) 
%&=&
  \int \cdots \int \frac{\mbox{d}^d k_1\ldots \mbox{d}^d k_l}
  {E_1^{a_1}\ldots E_n^{a_n}}\,,$
%  \label{FI}$
%\end{eqnarray}
where $d=4-2\ep$ is the space-time dimension, $a_n$ are
indices, $l$ is the number of loops and $1/E_n$ are propagators. We
work in Minkowski space where
the standard propagators are the form $1/(m^2 -p^2-i0)$.
Other propagators are permitted, see \cite{FIESTA}.
Substituting
%\begin{eqnarray}
$    \frac{1}{E_i^{a_i}}=\frac{e^{ai\pi/2}}{\Gamma(a)}\int_0^\infty \mbox{d}\alpha \alpha^{a_i-1} e^{-iE_i\alpha},$
%\label{AlphaSubstitution}
%\end{eqnarray}
after usual tricks \cite{FIESTA}, performing the decomposition of the integration region into the so-called
\textit{primary sectors} \cite{BH} and making a variable replacement, one results
in a linear combination of integrals
%\begin{eqnarray}
$\int_{x_j=0}^1 d x_i\ldots d x_{n'}\left(\prod_{j=1}^{n'} x_j^{a_j-1}\right) \frac{U^{A-(l+1)d/2}}{F^{A-ld/2}}$
%\label{Cube}
%\end{eqnarray}

If the functions $\frac{U^{A-(l+1)d/2}}{F^{A-ld/2}}$ had no
singularities in $\ep$, one would be able to perform the expansion in
$\ep$ and perform the numerical integration afterwards. However, in
general one has to resolve the singularities first.
%which is not possible for general $U$ and $F$. 
Thus, one starts a process the
sector decomposition aiming to end with a sum of similar expressions,
but with new functions $U$ and $F$ which have no singularities (all
the singularities are now due to the part $\prod_{j=1}^n
{x'}_j^{a'_j-1}$). The way sector decomposition is performed
is called a \textit{sector decomposition strategy} (\cite{BH,BognerWeinzierl,FIESTA}) and is an essential
part of the algorithm (%In \cite{BognerWeinzierl} four strategies were proposed; three of
%those are guaranteed to terminate while the last one is a heuristic
%strategy, which likely shares the ideas of \cite{BH}, is not
%guaranteed to terminate but results in less sectors than the other
%strategies. All these strategies are
%implemented in {\tt FIESTA}; by default
%it uses a new strategy based on original ideas \cite{FIESTA}.
let us also mention a geometrical
approach to sector decomposition \cite{Kaneko} which is rather
complicated in implementation as a strategy on a computer but promises
to be the optimal one).

After the sector decomposition one resolves the singularities by
evaluating the first terms of the Taylor series:
in those terms one integration is taken analytically.
%, and the remainder
%has no singularities. 
Afterwards the $\ep$-expansion can be performed
and finally one can do the numerical integration.

{\tt FIESTA} is written in {\tt Mathematica} \cite{math7}  and C.  The user
is not supposed to use the C part directly as it is launched from {\tt
  Mathematica} via the Mathlink protocol in order to perform a
numerical integration. 
%By default {\tt FIESTA} uses the {\tt Vegas} integrator but this
%behavior can be easily controlled by the user. Both Mathematica and C
%parts can be efficiently parallelized on modern multi-core computers; the C part
%also parallelizable on clusters.
%The {\tt FIESTA} user interface is based on {\tt
%Mathematica}. 
To run {\tt FIESTA}, the user has to load the {\tt
FIESTA} package into {\tt  Mathematica} 6 or 7. In order to evaluate a
Feynman integral one has to use the command
%\begin{alltt}
{\tt SDEvaluate[UF[loop\_momenta,propagators,subst],} {\tt indices,order]},
%\end{alltt}
where {\tt loop\_momenta} is a list of all loop momenta, {\tt
propagators} is a list of all propagators, {\tt subst} is a list of
substitutions for external momenta, masses and other values.
%(please
%note that the code performs numerical integrations, therefore the
%functions {\tt U} and {\tt F} should not depend on any external
%kinematic invariants), {\tt indices} is a list of indices and {\tt
%order} is a required order of the $\ep$-expansion.
For example,\\
%[1em]
\hspace*{1ex}{\tt SDEvaluate[UF[\{k\},\{-k$^2$,-(k+p$_1$)$^2$,-(k+p$_1$+p$_2$)$^2$,-(k+p$_1$+p$_2$+p$_4$)$^2$\},
\\
\hspace*{1ex}\{p$_1^2\rightarrow$0,p$_2^2\rightarrow$0,p$_4^2\rightarrow$0,
p$_1$ p$_2\rightarrow$-s/2,p$_2$ p$_4\rightarrow$-t/2,p$_1$ p$_4\rightarrow$-(s+t)/2,
\\
\hspace*{1ex}s$\rightarrow$-3,t$\rightarrow$-1\}],
\{1,1,1,1\},0]
}\\
%[1em]
evaluats the massless on-shell box diagram
with Mandelstam variables equal to $-3$ and $-1$.

\section{Numerical results for four-loop massless propagators}

\begin{figure}[!hbt]
\begin{center}
\SetScale{0.8}
\SetWidth{1.0}

\vspace{-1.2 cm}
\begin{picture}(75,90)(0,0)
%\put(40,80){m61}
\CArc(50,50)(20,0,360)
\Line(70,50)(80,50)
\Line(30,50)(20,50)
\Line(61,67)(61,33)
\Line(39,67)(39,34)
\Line(39,50)(61,50)
\put(33,8){$M_{61},\,\ep^1$}
\end{picture}
\begin{picture}(75,90)(0,0)
%\put(40,80){m62}
\CArc(50,50)(20,0,360)
\Line(70,50)(80,50)
\Line(30,50)(20,50)
\Line(62,34)(62,47)
\Line(38,34)(38,47)
\Line(38,47)(62,47)
\Line(62,47)(36,64.3333333333333)
\Line(38,47)(46.3205029433784,52.5470019622523)
\Line(64,64)(53,56.6666666666667)
\put(33,8){$M_{62},\,\ep^0$}
\end{picture}
\begin{picture}(75,90)(0,0)
%\put(40,80){m63}
\CArc(50,50)(20,0,360)
\Line(70,50)(80,50)
\Line(30,50)(20,50)
\Line(39,67)(39,33)
\Line(39,50)(68,57.25)
\Line(61,33)(61,53)
\Line(61,67)(61,58)
\put(33,8){$M_{63},\,\ep^0$}
\end{picture}
\begin{picture}(75,90)(0,0)
%\put(40,80){m51}
\CArc(50,50)(20,0,360)
\Line(70,50)(80,50)
\Line(30,50)(20,50)
\Line(50,70)(34,38)
\Line(40,50)(65,37.5)
\Line(50,70)(50,48)
\Line(50,30)(50,42)
\put(33,8){$M_{51},\,\ep^1$}
\end{picture}
%\vskip -2.em
\begin{picture}(75,90)(0,0)
%\put(40,80){m41}
\CArc(50,50)(20,0,360)
%\put(50,50){\line(2,3){11}}
%\put(50,50){\line(-2,-3){11}}
\Line(61,67)(52,53.5)
\Line(39,34)(48,47.5)
\Line(50,50)(39,66.5)
\Line(50,50)(61,33.5)
\Line(61,67)(39,67)
\Line(70,48)(80,48)
\Line(30,48)(20,48)
\put(33,8){$M_{41},\,\ep^1$}
\end{picture}

\vspace{-1.2 cm}
\begin{picture}(75,90)(0,0)
%\put(40,80){m42}
\CArc(50,50)(20,0,360)
%\put(50,50){\line(2,3){11}}
%\put(50,50){\line(-2,-3){11}}
\Line(61,66)(52,52.5)
\Line(39,34)(48,47.5)
\Line(50,50)(39,66.5)
\Line(50,50)(61,33.5)
\Line(61,66)(70,48)
\Line(70,48)(80,48)
\Line(30,48)(20,48)
\put(33,8){$M_{42},\,\ep^1$}
\end{picture}
\begin{picture}(75,90)(0,0)
%\put(40,80){m44}
\CArc(50,50)(20,0,360)
\Line(50,50)(61,66.5)
\Line(50,50)(39,33.5)
%\put(50,50){\line(-2,3){11}}
%\put(50,50){\line(2,-3){11}}
\Line(70,48)(80,48)
\Line(30,48)(20,48)
\Line(61,67)(61,33)
\Line(39,67)(39,34)
\put(33,8){$M_{44},\,\ep^0$}
\end{picture}
\begin{picture}(75,90)(0,0)
%\put(40,80){m45}
\CArc(50,50)(20,0,360)
%\put(50,50){\line(2,3){11}}
%\put(50,50){\line(-2,-3){11}}
\Line(61,66)(52,52.5)
\Line(39,34)(48,47.5)
\Line(50,50)(39,66.5)
\Line(50,50)(61,33.5)
\Line(61,67)(61,34)
\Line(70,48)(80,48)
\Line(30,48)(20,48)
\put(33,8){$M_{45},\,\ep^1$}
\end{picture}
%\vskip -2.0em
\begin{picture}(75,90)(0,0)
%\put(40,80){m34}
\Line(30,50)(20,50)
\Line(70,50)(80,50)
\CArc(50,50)(20,0,360)
\Line(34,62)(66,62)
\Line(50,30)(34,62)
\Line(50,30)(66,62)
\put(33,8){$M_{34},\,\ep^3$}
\end{picture}
\begin{picture}(75,90)(0,0)
%\put(40,80){m35}
\Line(30,50)(20,50)
\Line(50,50)(80,50)
\CArc(50,50)(20,0,360)
\Line(50,50)(41,68)
\Line(50,50)(41,32)
\Line(41,32)(41,68)
\put(33,8){$M_{35},\,\ep^2$}
\end{picture}

\vspace{-1.2 cm}
\begin{picture}(75,90)(0,0)
%\put(40,80){m36}
\Line(20,50)(80,50)
\CArc(50,50)(20,0,360)
\Line(50,30)(50,70)
\put(33,8){$M_{36},\,\ep^1$}
\end{picture}\begin{picture}(75,90)(0,0)
%\put(40,80){m52}
\CArc(40,50)(10,0,360)
\CArc(66,50)(15,0,360)
\Line(55,61)(77,39)
%\Line(77,61)(55,39)
\Line(77,61)(67.54,51.54)
\Line(64.46,48.46)(55,39)
\Line(30,50)(20,50)
\Line(82,50)(92,50)
\put(33,8){$M_{52},\,\ep^1$}
\end{picture}
\begin{picture}(75,90)(0,0)
%\put(40,80){m43}
\CArc(50,50)(20,0,360)
\Line(20,50)(80,50)
\Line(42,68)(60,50)
%\Line(58,68)(40,50)
\Line(58,68)(51.6496,61.6496)
\Line(47.56,57.56)(40,50)
\put(33,8){$M_{43},\,\ep^1$}
\end{picture}
\begin{picture}(75,80)(-15,-10)
\SetScale{1.0}
\SetWidth{0.8}
\put(15,0){$N_0,\ep^2$}
\CArc(26,30)(15,0,360)
\Line(15,41)(37,19)
\Line(37,41)(28,32)
\Line(15,19)(23.4852813742386,27.4852813742386)
\Line(10,30)(5,30)
\Line(42,30)(47,30)
\end{picture}
\caption{ $M_{61}$--$M_{43}$: the thirteen complicated four-loop master integrals according to \cite{BaCh}.
%The integrals are ordered  (if  read from left to right and then from top to bottom) according 
%to their complexity. 
The two MI's $M_{52}$ and $M_{43}$ can be identically  expressed through  the  three-loop nonplanar MI $N_0$.  
}
\label{m41-63}
\end{center}
\end{figure}
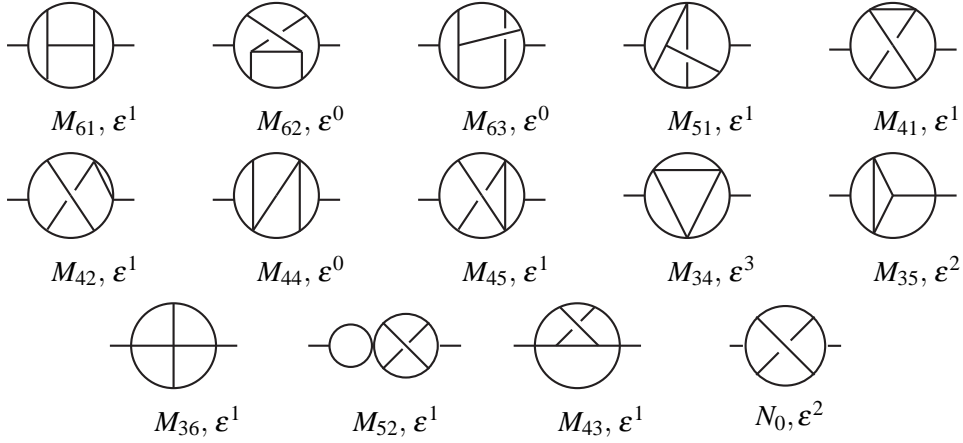

In \cite{BaikovCriterion} a full set of
four-loop massless propagator-like  MI's was identified.
There are 28 independent MI's. Analytical results for these
integrals were obtained in
\cite{BaCh}. 
The most complicated MI's are demonstrated on Fig.~\ref{m41-63}.
$\ep^{m}$ after $M_{ij}$ stands for the maximal term in
$\ep$-expansion of $M_{ij}$ which one needs to know for evaluation of
the contribution of the integral to the final result for a four-loop
integral after reduction is done, see \cite{BaCh}.
Two of the complicated integrals ($M_{43}$ and $M_{52}$) are related by
a simple factor with the three-loop MI $N_0$  \cite{propagators} so it is enough to evaluate remaining
eleven complicated MI's   $M_{61}$--$M_{36}$ as well as first three terms  of the $\ep$-expansion of $N_0$. 

We calculated them (for $q^2 = -1$) using {\tt FIESTA}
with the {\tt Cuba}\cite{Cuba} {\tt Vegas}
integrator and 1 500 000 sampling points for integration. 
%The square of
%the external momentum $q$ was chosen as -1: .
Our results alongside with the corresponding analytical expressions
(tran\-sformed to the numerical form) from \cite{BaCh} look like
follows\footnote{Please, note that the overall normalization used by
{\tt FIESTA} is {\em different} from the one employed by the authors
of \cite{BaCh}, see \cite{propagators}. }:

\begin{itemize}
\item[$\ovl{M}_{34}$] $\ep^{-4}$: 0.08333 $\pm$ 0 (0.08333);
  $\ep^{-3}$: 0.916667
  $\pm$ 0.000018 (0.91666); $\ep^{-2}$: 5.64251 $\pm$ 0.00022 (5.6425109); $\ep^{-1}$:
  27.6413 $\pm$ 0.00077 (27.6412581); $\ep^{0}$: 98.638 $\pm$ 0.0034
  (98.637928); $\ep^{1}$: 342.736 $\pm$ 0.012 (342.7349920); $\ep^{2}$:
  857.88 $\pm$ 0.048 (857.8735165); $\ep^{3}$: 2659.84 $\pm$ 0.19 (2659.825402); $\ep^{4}$:
  4344.28 $\pm$ 0.75 (unknown); $\ep^{5}$: 17483.1 $\pm$ 5.7 (unknown).
\item[$\ovl{M}_{35}$]$\ep^{-2}$: 0.601028 $\pm$ 0.000012
  (0.601028); $\ep^{-1}$: 7.4231 $\pm$ 0.00024 (7.423055); $\ep^{0}$:
  44.9127 $\pm$ 0.00073 (44.91255); $\ep^{1}$: 217.023 $\pm$ 0.0037
  (217.0209); $\ep^{2}$: 780.436 $\pm$ 0.013 (780.432);
  $\ep^{3}$: 2678.13 $\pm$ 0.053 (unknown); $\ep^{4}$: 7195.9 $\pm$
  0.3 (unknown).
\item[$\ovl{M}_{36}$] $\ep^{-1}$: 5.184645 $\pm$ 0.000042
  (5.1846388); $\ep^{0}$: 38.8948 $\pm$ 0.00039 (38.8946741);
  $\ep^{1}$: 240.069 $\pm$ 0.0019 (240.0684359); $\ep^{2}$: 948.623
  $\pm$ 0.0091 (unknown); $\ep^{3}$: 3679.77$\pm$ 0.06 (unknown).
\item[$\ovl{M}_{41}$]$\ep^{-1}$: 20.73860 $\pm$
  0.00023(20.7385551); $\ep^{0}$: 102.033 $\pm$ 0.003 (102.0326759);
  $\ep^{1}$: 761.60 $\pm$ 0.011 (761.5969858); $\ep^{2}$: 2326.18
  $\pm$ 0.062 (unknown); $\ep^{3}$: 12273.6 $\pm$ 0.4 (unknown).
\item[$\ovl{M}_{42}$]$\ep^{-1}$: 20.73860 $\pm$ 0.00024 (20.7385551);
  $\ep^{0}$: 145.381 $\pm$ 0.0029 (145.3808999); $\ep^{1}$: 985.91
  $\pm$ 0.014 (985.9082306); $\ep^{2}$: 3930.65 $\pm$ 0.076 (unknown);
   $\ep^{3}$: 17486.6 $\pm$ 0.6 (unknown).
\item[$\ovl{M}_{44}$] $\ep^{0}$: 55.58537 $\pm$ 0.00031 (55.5852539);
  $\ep^{1}$: 175.325 $\pm$ 0.004 (unknown); $\ep^{2}$: 1496.52 $\pm$ 0.02
\item[$\ovl{M}_{45}$] $\ep^{0}$: 52.0181 $\pm$ 0.0003
  (52.0178687); $\ep^{1}$: 175.50 $\pm$ 0.0036 (175.496447); $\ep^{2}$:
  1475.272 $\pm$ 0.0098 (unknown); $\ep^{3}$: 2623.5 $\pm$ 0.1 (unknown).
\item[$\ovl{M}_{51}$] $\ep^{-1}$: -5.184651 $\pm$ 0.000048
  (-5.184638); $\ep^{0}$: -32.0962 $\pm$ 0.00057 (-32.09614);
  $\ep^{1}$: -91.158 $\pm$ 0.0052 (-91.1614); $\ep^{2}$: 119.06 $\pm$
  0.043 (unknown); $\ep^{3}$: 2768.6 $\pm$ 0.45 (unknown).
\item[$\ovl{N}_0$] $\ep^{0}$: 20.73857 $\pm$ 0.00026 (20.7385551);
  $\ep^{1}$: 190.60 $\pm$ 0.0023 (190.600238);  $\ep^{2}$: 1049.20
  $\pm$ 0.014 (1049.194196); $\ep^{3}$: 4423.84 $\pm$ 0.072 (unknown);
  $\ep^{4}$: 16028.8 $\pm$ 0.5 (unknown).
\item[$\ovl{M}_{61}$] $\ep^{-1}$: -10.36931 $\pm$ 0.00006
  (-10.3692776); $\ep^{0}$: -70.990 $\pm$ 0.0011 (-70.99081719);
  $\ep^{1}$: -21.650 $\pm$ 0.013 (-21.663005); $\ep^{2}$: 2832.69
  $\pm$ 0.096 (unknown).
\item[$\ovl{M}_{62}$] $\ep^{-1}$: -10.36933 $\pm$ 0.00006
  (-10.36927); $\ep^{0}$: -58.6187 $\pm$ 0.0013(-58.6210);
  $\ep^{1}$: 244.681 $\pm$ 0.015 (unknown).
\item[$\ovl{M}_{63}$] $\ep^{-1}$: -5.18467 $\pm$ 0.000042
  (-5.184638); $\ep^{0}$: 14.3989 $\pm$ 0.00081 (14.39739);
  $\ep^{1}$:  739.979 $\pm$ 0.0099 (unknown).
\end{itemize}
Here for each  MI we provide our numerical result for coefficients of
$\ep$-expansion in comparison (in parentheses) with the known from \cite{BaCh}
analiycal results (if any). As we can see, our calculations reproduce the
result of \cite{BaCh} with 3-4 correct digits.
The extra terms in the $\ep$-expansion
of each MI which are currently unavailable analytically but are
necessary for future five-loop calculations.

\section{Conclusion}

Usually, analytical evaluation of multiloop MI is a kind of
art. It requires a lot of efforts (and CPU
time). In many situations, independent checkup is hardly any possible
in reasonable time. That is why the simple in use tools for numerical
evaluation like {\tt FIESTA} are  important.

\vspace{0.2 cm}

{\em Acknowledgments.}
This work was supported in part by DFG through SBF/TR 9 and the
Russian Foundation for Basic Research through grant 08-02-01451.

\end{document}